\documentclass[%
 reprint,
 amsmath,amssymb,
 aps,
]{revtex4-2}

\usepackage{color}
\usepackage{graphicx}% Include figure files
\usepackage{dcolumn}% Align table columns on decimal point
\usepackage{bm}% bold math
\usepackage{hyperref}% add hypertext capabilities
\usepackage[mathlines]{lineno}% Enable numbering of text and display math
\begin{document}
% \begin{CJK}{UTF8}{gbsn}

\preprint{APS/123-QED}

\title{The Thermodynamic Model to Study the Slow Afterhyperpolarization in a Single Neuron at Different ATP Levels}

\author{Jianwei Li$^1$}
 \email{These authors contribute equally to this work.}
\author{Simeng Yu$^1$$^*$}
\author{Mingye Guo$^2$$^*$}
\author{Xuewen Shen$^1$}
\author{Qi Ouyang$^3$}
\author{Fangting Li$^1$}%
 \email{Corresponding author: lft@pku.edu.cn}
\affiliation{%
 $^1$School of Physics, Peking University, Beijing 100871, China
}%
\affiliation{%
 $^2$School of Basic Medical Sciences, Peking University Health Science Center, Beijing, 100191, China
}%
\affiliation{%
 $^3$Department of Physics, Zhejiang University, Hangzhou 310027, Zhejiang Province, China
}%
% \author{Jianwei Li, Simeng Yu, Mingye Guo, Xuewen Shen, Fangting Li}
% \author{Jianwei Li\and Simeng Yu\and Mingye Guo\and Xuewen Shen\and Fangting Li}

% %% Authors （在这用不了）
% \author[a]{Jianwei Li\footnote{These authors contribute equally to this work }}
% \author[a]{Simeng Yu$^*$}
% \author[a]{Mingye Guo$^*$}
% \author[a]{Xuewen Shen}
% \author[a]{Fangting Li \thanks{Corresponding author: lft@pku.edu.cn}}
% \affil[a]{School of Physics, Peking University, Beijing 100871, China }
% \affil[b]{School of Physics, Peking University, Beijing 100871, China }

\date{\today}

\begin{abstract}
The neuron consumes energy from ATP hydrolysis to maintain a far-from-equilibrium steady state inside the cell, thus all physiological functions inside the cell are modulated by thermodynamics.  The neurons that manage information encoding, transferring, and processing with high energy consumption, displaying a phenomenon called slow afterhyperpolarization after burst firing, whose properties are affected by the energy conditions. Here we constructed a thermodynamical model to quantitatively describe the sAHP process generated by $Na^+-K^+$ ATPases(NKA) and the Calcium-activated potassium(K(Ca)) channels. The model simulates how the amplitude of sAHP is effected by the intracellular ATP concentration and ATP hydrolysis free energy $\Delta$ G. The results show a trade-off between NKA and the K(Ca)'s modulation on the sAHP's energy dependence, and also predict an alteration of sAHP's behavior under insufficient ATP supply if the proportion of NKA and K(Ca)'s expression quantities is changed. The research provides insights  in understanding the maintenance of neural homeostasis  and support furthur researches on  metabolism-related and neurodegenerative diseases.
\end{abstract}

%\keywords{Suggested keywords}%Use showkeys class option if keyword
                              %display desired
\maketitle

%\tableofcontents
%\added{1.Try $\backslash$include and $\backslash$input commands, which would help you devide the main text into several independent parts with multi tex files.}\\

%\added{2.Try $\backslash$mathline.lineno to mark lines in main text for the convience of modification.}
\section{\label{sec:level1}Introduction}
The living cells are open systems far from equilibrium. The cell membrane's selectivity keeps the independence of cell and meanwhile allows energy input, which is the foundation of non-equilibrium steady state. At the center of all intracellular biochemical reactions stands the ATP hydrolysis reaction, a universal energy supplying reaction in the cell. Most of the cellular physiological functions have to couple with ATP hydrolysis to maintain steady reaction flux\cite{Qian_2021}.\par
Single neural coding, considered the basis of information processing in human brain, is effected significantly by ATP hydrolysis. Neural coding is energetically expensive: actually the brain only takes up 2\% of human body mass but cost 20\% of the energy supply\cite{Ma_2017}. The fact is that neural spiking itself does not directly involve energy consumption, since action potential is generated by passive transportation of voltage-gated sodium and potassium channels, which is quantitatively modelled by Hodgkin and Huxley in 1952\cite{Hodgkin_1952}. However, the neuron  uses mainly the $Na^+$-$K^+$ ATPase powered by ATP hydrolysis to maintain stable ion concentration gradients across the membrane. \par
So it is worth discussing whether the neuron’s behavior will be influenced when its energy supply is altered . To be more specific, we focused on a process that has high demand of energy, that is the bursting. Neural bursting is periods of rapid action potential spiking  followed by quiescent periods much longer than typical inter-spike intervals, a diverse firing pattern in central nervous system. A burst typically last hundreds of milliseconds or even seconds, causing distinct variation of neural ion concentration, which results in a high demand of $Na^+$-$K^+$ ATPase turnover rate. The $Na^+$-$K^+$ ATPase continues to work at high speed for some time after a burst and generate a long lasting hyperpolarization, which is termed the slow afterhyperpolarization(sAHP).\par
Beside the $Na^+$-$K^+$ ATPase, the Calcium-activated-potassium channel also can  generate sAHP, which involves another two ATP-hydrolysis ATPases. Actually it is considered the only origin of sAHP when this phenomenon is discovered in 1980s\cite{Alger_1980}. While the subsequent researches shows that $Na^+$-$K^+$ ATPase also plays an important role in the generation of sAHP in CA1 pyramidal neurons\cite{Gulledge_2013}, lamprey sensory neurons\cite{Parker_1996}, presynaptic and postsynaptic Calyx of Held neurons\cite{Kim_2007}, dopaminergic midbrain neurons\cite{Johnson_1992}, etc. Tiwari et al.\cite{Tiwari_2018} gives a relatively complete comparison about K(Ca) channel and $Na^+$-$K^+$ ATPases' contribution in generating sAHP and pointed out their distinctions such as the different decaying time scale.\par
sAHP is found important in suppressing neural excitability\cite{Arganda_2007}\cite{Zhang_2012}, this property is observed to act as an negative feed back in Xenopus frog tadpoles' spinal locomotor network to keep the stability of the network\cite{Zhang_2015}, there are also evidence shows that sAHP prevent neuron sliding into an epileptic pattern\cite{Alger_1980}. Some researches also state that sAHP displays characteristics of memory \cite{Forrest_2014}.\par
However, there are not many discussion about sAHP's energy dependence although its dual generating mechanisms are both closely linked with ATP level and ATP hydrolysis free energy in the neuron. Here we constructed a biophysically-based model to quantitatively describe the two sAHP generating mechanisms discussed above. Based on the model, we are able to discuss how will the intracellular energy supply condition affects the sAHP and in turn has an effect on the neural behavior after the burst. To verify the above results, we searched for relevant data on mouse neurons and found that the behaviour of hippocampal neurons and the water maze might be related. We also carried out relevant simulation studies and theoretical predictions using the model.

\section{\label{sec:level2}Models}
The Hodgkin-Huxley model\cite{Hodgkin_1952}, as a mathematical model that effectively describes neuronal action potentials, provides a model-based paradigm for studying neurons. Based on the Hodgkin-Huxley model and extended to include other kinds of ion channels, it is an appropriate tool for studying the AHP process. So far, there are already some single compartment models that focus on the mechanism of sAHP, including the modulation of $Na^+$-$K^+$ ATPases or Calcium-activated-potassium channels alone, like models raised by Burrell et al\cite{Burrell_2008}, Kudela et al.\cite{Kudela_2009}, Lee et al.\cite{Lee_2010}; others systematically describes the two pathways, like the works of Csercsik et al.\cite{Csercsik_2011}, Chen et al.\cite{Chen_2013} and Moran et al.\cite{Moran_2016}. There also some multi-compartment or PDE models that pay more attentions on the combination of sAHP transition and neural morphology, like the works of Cataldo et al.\cite{Cataldo_2005}, or sAHP's modulation on the interactions between different part of the neuron or between different neurons, like the works of Krishnan et al.\cite{Krishnan_2015}. Comparing with the previous work, our single compartment model puts emphasis on the coupling of ATP hydrolysis with the two kinds of sAHP generating mechanism by applying thermodynamic models of ATPases, and explores how the two mechanisms interact with each other.\par

FIG \ref{fig:1}a. illustrates the model components and their interactions. Based on two pathways to generate sAHP, the model is divided into a sub-model for membrane voltage changes and a sub-model  for $Ca^{2+}$ concentration changes, and they are linked by $Ca^{2+}$'s activation of K(Ca) channel, which is plotted as a dotted arrow in the graph. The main modeling assumptions and equations are listed below. A more detailed discussion can be found in Appendix \ref{app:model}.

\subsection{sub-model for membrane voltage changes}
This sub-model is based on a single-compartment Hodgkin-Huxley framework\cite{Hodgkin_1952}, and applied the modification by Traub et al.\cite{Traub_1991} to fit the data of mouse's pyramidal neuron which is observed to have sAHP phenomenon\cite{Gulledge_2013}. Changes of membrane potential over time is described by the ODE equation \ref{eq:1}.
\begin{equation}
    \frac{\mathrm{d}V}{\mathrm{d}t}=-\frac{1}{C_m}(I_{inj}+I_{Na(V)}+I_{K(V)}+I_{K(Ca)}+I_{NKA}+I_{leak})
    \label{eq:1}
\end{equation}
Here V denotes the membrane potential (mV), $C_m$ denotes the membrane's capacitance per unit area ($\mu \rm F/mm^2$), and $I_x$ with different subscripts $x$ denotes membrane current density (per unit area, $\mu \rm A/mm^2$, setting the direction of flowing out of the neuron to be positive) through different channels, pumps and electrode, respectively.
\paragraph{$I_{inj}$} is the simulation for the current injected into the neuron through an electrode to start neural spiking.
\paragraph{$I_{Na(V)}$} is the current through voltage-gated sodium channels. By applying Hodgkin-Huxley model:
\begin{equation}\label{eq:2}
   I_{Na(V)}=g_{Na}m^3h(V-E_{Na})
\end{equation}
Here $g_{Na}$ is the maximum conductance of $Na(V)$ channels per unit area, m and h are gating variables controlled by membrane voltage, and $E_{Na}$ is the reversal potential of $Na^+$ calculated by the Nernst equation:
\begin{equation*}
    E_{Na}=\frac{RT}{F}\mathrm{ln}(\frac{[Na]_e}{[Na]_i})
\end{equation*}
$[Na]_e$ and $[Na]_i$ are the extracellular and intracellular concentrations of the sodium ions.

\paragraph{$I_{K(V)}$} is the current through voltage-gated potassium channels. Similarly, by using $g_{K}$ to denote the maximum conductance of K(V) channels per unit area and $n$ to denote the gating variable:
\begin{equation}\label{eq:10}
   I_{K(V)}=g_{K}n^4(V-E_{K})
\end{equation}
with reversal potential $E_K$
\begin{equation*}
    E_{K}=\frac{RT}{F}\mathrm{ln}(\frac{[K]_e}{[K]_i})
\end{equation*}

\paragraph{$I_{K(Ca)}$}
$I_{K(Ca)}$ is the current through calcium-activated potassium channels (also called the small conductance potassium channels, SK channels), the component directly involving in the sAHP process. This type of potassium channel opens with calcium binding on. Based on the K(Ca) channel model described in \cite{Engel_1999}, $I_{K(Ca)}$ can be written as \eqref{eq:15}:
\begin{equation}\label{eq:15}
   I_{K(Ca)}=g_{K(Ca)}\cdot \omega (Ca)\cdot (V-E_{K})
\end{equation}
Here $g_{K(Ca)}$ is the maximum conductance of K(Ca) channels per unit area, and $\omega(Ca)$ is the Michaelis–Menten equation with Hill coefficient = 2 to simulate the opening probability related with calcium binding:
\begin{equation*}
   \omega (Ca)=\frac{[Ca]^2_i}{[Ca]^2_i+k_{Ca}^2}
\end{equation*}

\begin{figure*}
\centering
\includegraphics[width=1\linewidth]{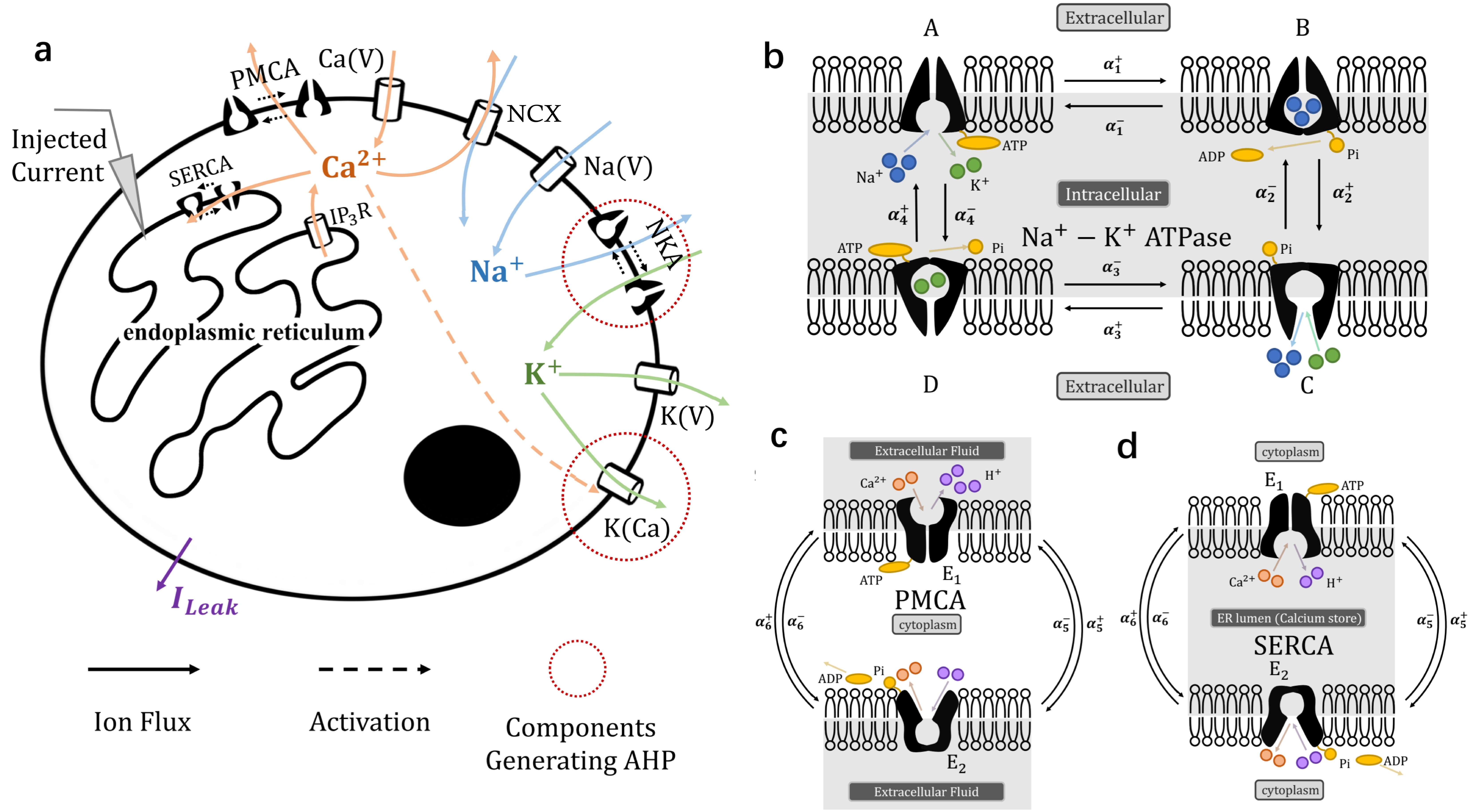}% Here is how to import EPS art
\caption{\label{fig:1} \textbf{The schematic representation for modeling slow afterhyperpolarization(sAHP) generation process in a single neuron.}\ \textbf{a.}Illustration of all the pathways for $Na^+$, $K^+$ and $Ca^{2+}$ ionic flows and their interactions in the model. The dotted arrow shows $Ca^{2+}$'s activation of the K(Ca) channels.\ \textbf{b.}$Na^+$-$K^+$ ATPase’s kinetic mechanism. For every single ATP consumed, the ATPase pumps 3 $Na^+$ out of and 2 $K^+$ into the cell. \ \textbf{c.}The mechanism of Plasma membrane $Ca^{2+}$ ATPase (PMCA) and sarco(endo)plasmic reticulum $Ca^{2+}$ ATPase (SERCA). Belonging to the P-type ATPase family as well, $Ca^{2+}$ ATPase has a similar behavior to the $Na^+$-$K^+$ ATPase. According to \cite{Pan_2019}, its mechanism can be further simplified to a 2-state cycle, as illustrated on the diagram. Different from $Na^+$-$K^+$ ATPase, both PMCA and SERCA's behaviors are not affected by the membrane potential, since there is no net current through the PMCA and no potential difference across the SERCA. \\
    $^{\textbf{*}}$ Meanings of the labels in \textbf{a.} \textbf{Na(V)}: voltage-gated $Na^+$ channel; \textbf{K(V)}: voltage-gated $K^+$ channel; \textbf{K(Ca)}: $Ca^{2+}$ activated $K^+$ channel; \textbf{NKA}: $Na^+$-$K^+$ ATPase; \textbf{Ca(V)}: voltage-gated $Ca^{2+}$ channel; \textbf{NCX}: sodium-calcium exchange; \textbf{PMCA}: Plasma membrane $Ca^{2+}$ ATPase; \textbf{IP$_3$R}: InsP$_3$-gated calcium channels on the endoplasmic reticulum membrane; \textbf{SERCA}: sarco(endo)plasmic reticulum $Ca^{2+}$ ATPase; $I_{Leak}$: leaky current; \textbf{Injected Current}: current injected to the neural soma by an electrode to stimulate neural spiking.}
\end{figure*}

\paragraph{$I_{NKA}$} is the current passing through $Na^+-K^+$ ATPases. This ATP-hydrolysis-driven pump consumes the free energy from 1 ATP molecule's hydrolysis to actively transport 3 $Na^+$ and 2 $K^+$ against their electrochemical gradients in a turnover cycle, causing an outward current of one positive charge per cycle. Therefore, $I_{NKA}$ can be written as (\ref{eq:17}):
\begin{equation}\label{eq:17}
   I_{NKA}=e\sigma_{NKA}J_{NKA}
\end{equation}
Here $J_{NKA}$ denotes the average turnover rate of $Na^+-K^+$ ATPases, and $\sigma_{NKA}$ denotes the surface density of $Na^+-K^+$ ATPases on the membrane.\par
A biophysically-based kinetic model underlying the energy transduction linked with ATP hydrolysis is applied in order to investigate thermodynamic properties of the sAHP. It is modified from Pan et al.'s work\cite{Pan_2019}, which is an simplification of the Post-Albers 15 states model\cite{Albers_1967}\cite{Post_1969}. Pan et al. proved that the 15 states model can be reduced to 4 states by assuming quasi-equilibrium for ion binding and unbinding process. The schematic mechanism for this reduced model is plotted as FIG \ref{fig:1}b. The 4 states are marked as A, B, C, D and the forward and backward transition rates between them are denoted as $\alpha_i^{\pm}$. State A and state C are two different conformations that face to the intracellular and extracellular side, respectively, while B and D are two occluded states during the conformational transitions triggered by ATP hydrolysis. State A has higher affinity for $Na^+$ than $K^+$, while state C is just the opposite: This property helps state A bind 3 cytosolic $Na^+$ and release 2 $K^+$ into the cytoplasm, while state C exchanges 2 $K^+$ for 3 $Na^+$ from the extracellular side\cite{Kaplan_2002}. \par
The mechanism for $Na^+$ binding is proved to be voltage-sensitive, to be more specific, the voltage dependence is mainly associated with only one $Na^+$\cite{Hilgemann_1994}. So the model assumes two $Na^+$ binding to the ATPase with voltage-independent dissociation constant $K_{d,Na_{i}}$ and $K_{d,Na_{e}}$, for the intracellular and extracellular sides, respectively; while the other $Na^+$ has the dissociation constant $K_{d,Na_{i0}}e^{\Delta V/RT}$ and $K_{d,Na_{e0}}e^{(1+\Delta) V/RT}$, here $K_{d,Na_{i0}}$ and $K_{d,Na_{e0}}$ are the voltage independent parts. The binding of other ions and molecules are all regarded as voltage-insensitive. \par
Thus the expression of $\alpha_i^{\pm}$ can then be deduced from the dissociation constants $K_{d,X_{i/e}}$, the ion concentrations $[X]_{i/e}$ (X can be Na, K or Ca) and the transition rates $k_i^{\pm}$ in the original 15 states model. These equations are attached in the Appendix.\par
A single $Na^+-K^+$ ATPase jumps between the 4 states through a Continuous-time Markov chain. Since it has a high membranal surface density, we denote the probability for the $Na^+-K^+$ ATPase being in state A, B, C, D as $P_A$, $P_B$, $P_C$, $P_D$ (with $P_A+P_B+P_C+P_D=1$) and use the law of mass action:
\begin{widetext}
\begin{equation}\label{eq:33}
\frac{\mathrm{d}}{\mathrm{d}t}
\left[
\begin{array}{c}
  P_A \\
P_B\\
   P_C\\
   P_D
\end{array}
\right]
=
\left[
\begin{array}{cccc}
-\alpha_1^+-\alpha_4^-& \alpha_1^- & 0& \alpha_4^+\\
\alpha_1^+ &-\alpha_2^+ -\alpha_1^-& \alpha_2^- & 0\\
0&\alpha_2^+& -\alpha_3^+-\alpha_2^-& \alpha_3^-\\
\alpha_4^-&0&\alpha_3^+&-\alpha_4^+-\alpha_3^-
\end{array}
\right]
\left[
\begin{array}{c}
  P_A \\
   P_B\\
  P_C\\
  P_D
\end{array}
\right]
\end{equation}
\end{widetext}

With fixed environmental conditions, the steady probability distribution can be calculated by assuming $\frac{\mathrm{d}P_A}{\mathrm{d}t}=\frac{\mathrm{d}P_B}{\mathrm{d}t}=\frac{\mathrm{d}P_C}{\mathrm{d}t}=\frac{\mathrm{d}P_D}{\mathrm{d}t}=0$ and solving the matrix equation. The free energy of ATP hydrolysis drives a steady reacting flow $J_{NKA}=\alpha_1^+A-\alpha_1^-B=\alpha_2^+B-\alpha_2^-C=\alpha_3^+C-\alpha_3^-D=\alpha_4^+D-\alpha_4^-A$. Actually this is just the average turnover rate of all $Na^+-K^+$ ATPases. But here we applied the model to the burst period, where the fluctuation of the membrane potential has similar time scale with most of the transition rates $\alpha_i^{\pm}$. The quasi-equilibrium assumption does not apply any more, so instead we solve ODE equations and set $J_{NKA}=\alpha_4^+D-\alpha_4^-A$, which is the flux related with intracellular ion exchanges.\par
To reach thermodynamical consistency, the model has to be in detailed balance when the Gibbs free energy $\Delta G=0$. It puts an restriction on the parameters :\begin{equation}\label{eq:48}
\frac{k_1^+k_2^+k_3^+k_4^+K^2_{d,K_i}K_{d0,Na_e}K^2_{d,Na_e}}{k_1^-k_2^-k_3^-k_4^-K_{d,ATP}K^2_{d,K_e}K_{d0,Na_i}K^2_{d,Na_i}}=K_{ATP, hyd}
\end{equation}
The derivation of \ref{eq:48} can be referred to in the Appendix\ref{app:model}.
We have kept this constraint when we adjust part of the parameters in the original model\cite{Pan_2019} to fit previous experiment results. Further details about parameter refitting can also be found in the Appendix\ref{app:model}.

\paragraph{$I_{leak}$} is the leaky current:
\begin{equation}\label{eq:49}
   I_{leak}=g_{L}(V-E_{L})
\end{equation}

\paragraph{The change of $[Na^+]_i$, $[K^+]_i$ concentrations}
The original Hodgkin-Huxley framework assumes constant $E_{Na}$ and $E_{K}$, because $[Na^+]_i$ and $[K^+]_i$ changes little in a single spike. But the process we discussed here involves long period burst, so the concentration changes of $Na^+$ and $K^+$ cannot be ignored anymore. The change of $[Na^+]_i$ and $[K^+]_i$ over time can be written as:
\begin{equation}\label{eq:50}
     \frac{\mathrm{d}[Na^+]_i}{\mathrm{d}t}=-\gamma_{1}(I_{Na(V)}/e+3\sigma_{NKA}J_{NKA})
\end{equation}
\begin{equation}\label{eq:51}
     \frac{\mathrm{d}[K^+]_i}{\mathrm{d}t}=-\gamma_{1}[ (I_{K(V)}+I_{K(Ca)})/e-2\sigma_{NKA}J_{NKA}]
\end{equation}
$\gamma_{1}$ is the volume-to-surface ratio when considering fluxes to the cytoplasm through the neuron membrane.\par
\eqref{eq:50} and \eqref{eq:51} also elucidate the basic physiological function of the $Na^+-K^+$ ATPase in the neuron: it helps to maintain the neuron's excitability. Without the $J_{NKA}$ term, $[Na^+]_i$ monotonously increases and $[K^+]_i$ decreases until $[Na^+]_i=[Na^+]_e$, $[K^+]_i=[K^+]_e$, where $E_{Na}=E_{K}=0$ and the neuron lose excitability.

\subsection{sub-model for calcium concentration changes}
$Ca^{2+}$ is crucial in lots of neural activities, and it associates with the sAHP through the activation of K(Ca) channel. In order to discuss the thermodynamical property of the hyperpolarization generated by the K(Ca) channel, we have to add the description about calcium concentration changes in our model. Different from $[Na^+]_i$ and $[K^+]_i$ that are of several or hundreds of mM, $[Ca^{2+}]_i$ is only $\sim 0.1 \mu$M in the resting state and the calcium current is also much smaller than the sodium and potassium current. So we have ignored calcium currents' contribution on the voltage variation. Another difference is that the cytosolic calcium can be replenished not only by the extracellular fluid, but also by the calcium store inside the neuron\cite{Verkhratsky_2005}. So here we applied a three-compartment model: the extracellular fluid, the cytoplasm and the endoplasmic reticulum (the calcium store). We used $[Ca^{2+}]_e$, $[Ca^{2+}]_i$, $[Ca^{2+}]_{er}$ to represent $Ca^{2+}$ concentrations in these compartments, respectively. The exchange pathways for calcium in the extracellular fluid and the cytoplasm are mainly the voltage-gated calcium channel (Ca(V)), sodium-calcium exchange (NCX) and the plasma membrane $Ca^{2+}$ ATPase (PMCA). While $Ca^{2+}$ can flow through the endoplasmic reticulum membrane mainly through the inositol 1,4,5-triphosphate gated calcium channels (IP$_3$R) and the sarco/endoplasmic reticulum $Ca^{2+}$ ATPase (SERCA). These pathways have been plotted in FIG \ref{fig:1}a. $[Ca^{2+}]_e$ is considered constant due to the large volume of the extracellular fluid, while the change of $[Ca^{2+}]_i$ and $[Ca^{2+}]_{er}$ over time can be written as:
\begin{eqnarray}\label{eq:52}
     \frac{\mathrm{d}[Ca^{2+}]_i}{\mathrm{d}t}&&=-\gamma_{1}((I_{Ca(V)}-2I_{NCX})/2e+\sigma_{PMCA}J_{PMCA})\nonumber\\
     &&-\gamma_{2}(\sigma_{SERCA}J_{SERCA}-J_{IP_3R})
\end{eqnarray}
\begin{equation}\label{eq:53}
     \frac{\mathrm{d}[Ca^{2+}]_{er}}{\mathrm{d}t}=\gamma_{3} (\sigma_{SERCA}J_{SERCA}-J_{IP_3R})
\end{equation}
Here $\gamma_{2}$ is the volume-to-surface ratio of fluxes flowing into the cytoplasm through the ER membrane, while $\gamma_{3}$ is the volume-to-surface ratio of fluxes into the ER lumen through the ER membrane. The reason for subtracting twice the $I_{NCX}$ here is that the net current of the $I_{NCX}$ is in the opposite direction of the $Ca$ current through the NCX, and, in addition, because a NCX carries three $Na$ ions and carries one $Ca$ ion, the magnitude of the $Ca$ current should be twice that of the $I_{NCX}$.

\paragraph{$I_{Ca(V)}$} is the current through the voltage-gated calcium channels on the neuron membrane, we also modeled it under Hodgkin-Huxley framework:
\begin{equation}\label{eq:54}
   I_{Ca(V)}=g_{Ca}p^2q(V-E_{Ca})
\end{equation}
and
\begin{equation*}
E_{Ca}=\frac{RT}{2F}\mathrm{ln}(\frac{[Ca^{2+}]_e}{[Ca^{2+}]_i})
\end{equation*}
$p$ and $q$ are voltage gating variables.

\paragraph{$I_{NCX}$} is the current through the sodium-calcium exchange on the neuron membrane. We adopted the simplest form of the equation that can describe the flip-flop of sodium-calcium exchange, from Mullins\cite{mullins1977mechanism}\cite{doi:10.1098/rstb.1985.0001}, as follows:
\begin{equation}\label{eq:54}
  I_{NCX}=\frac{k_{NCX}}{2}\left( \exp \frac{V-E_{NCX}}{RT/F}-\exp \frac{E_{NCX}-V}{RT/F} \right)
\end{equation}
and
\begin{equation*}
E_{NCX}=3E_{Na}-2E_{Ca}
\end{equation*}
$E_{NCX}$ is the reversal potential of NCX. If the membrane potential negative than $E_{NCX}$ , NCX acts in the forward-mode, exchanging sodium ions into the cell and calcium ions out of the cell; if the membrane potential positive than $E_{NCX}$, NCX acts in the reverse-mode, and exchanges reverse direction.

\paragraph{$J_{IP_3R}$} represents the calcium flux out of the ER, controlled by the intracellular messenger inositol 1,4,5-triphosphate (IP$_3$). The concentration of IP$_3$ can be considered constant in the sAHP process, but the opening probability of this channel can also be modulated by $[Ca^{2+}]_i$. Experiments have pointed out that calcium's modulation shows a biphasic behavior: the protein has both $Ca^{2+}$-binding activation domain and $Ca^{2+}$-binding inactivation domain facing the cytoplasm, so the opening probability first increases and then decreases as $[Ca^{2+}]_i$ rises. Based on De Young-Keizer model\cite{De_Young_1992}\cite{Csercsik_2011} that quantitatively describes this process, $J_{IP_3R}$ takes the form:
\begin{widetext}
\begin{equation}\label{eq:60}
J_{IP_3R}=(K_f(y(\frac{[Ca^{2+}]_i}{[Ca^{2+}]_i+K_a})(\frac{[IP_3]}{[IP_3]+K_i}))^3+J_{er})([Ca^{2+}]_{er}-[Ca^{2+}]_i)
\end{equation}
\end{widetext}
Here $\frac{[IP_3]}{[IP_3]+K_i}$ describes IP$_3$ binding, $\frac{[Ca^{2+}]_i}{[Ca^{2+}]_i+K_a}$ describes $Ca^{2+}$ binding to the activation domain, while $y$ represents the ratio of free $Ca^{2+}$-binding inactivation domain. The time variation of $y$ can be written as:
\begin{equation*}
   \frac{\mathrm{d}y}{\mathrm{d}t}=A(K_d(1-y)-y[Ca^{2+}]_i)
\end{equation*}
$K_d$ is the dissociation constant of $Ca^{2+}$ binding to the inactivation domain.

\paragraph{$J_{PMCA}/J_{SERCA}$} are $[Ca^{2+}]$ flux through PMCA and SERCA. These ATPases also consume energy from ATP hydrolysis to actively transport $Ca^{2+}$ and $H^{+}$, like the NKA does for $Na^+$ and $K^+$. The schematic mechanism for these two ATPases are plotted in FIG \ref{fig:1}c. Based on Inesi’s work\cite{Inesi_1987}, the overall reaction equations can be simplified as:
\begin{equation*}
2Ca_i^{2+}+E_1+ATP\rightleftharpoons E_2+ADP
\end{equation*}
\begin{equation*}
E_2 \rightleftharpoons 2Ca_e^{2+}+E_1+Pi
\end{equation*}
$E_1$ represents the state facing the side with higher calcium ion concentration (the extracellular fluid and the ER lumen) and $E_2$ represents the state facing the cytoplasm.\par
Here we left out the $[H^+]_i$ part since the three compartments are all able to maintain their pH value through $H^+$ buffer. We took $[H^+]$ as constants and integrated it in the model parameters.\par
Then the differential equations for probabilities $P_{E_1}$ and $P_{E_2}$ of PMCA are:
\begin{eqnarray}\label{eq:64}
\frac{\mathrm{d}P_{E_1}}{\mathrm{d}t}&&=-k_5^+P_{E_1}[Ca^{2+}]_i^2[ATP]_i+k_5^-P_{E_2}[ADP]_i\nonumber\\
&&+k_6^+P_{E_2}-k_6^-P_{E_1}[Ca^{2+}]_e^2[Pi]_i
\end{eqnarray}
\begin{equation}\label{eq:65}
\frac{\mathrm{d}P_{E_2}}{\mathrm{d}t}=-\frac{\mathrm{d}P_{E_1}}{\mathrm{d}t}
\end{equation}
The differential equations for SERCA can be derived similarly.\par
Different from the NKA that causes an outward ion current when working, PMCA transports 2 $Ca^{2+}$ from cytoplasm to the extracellular fluid to exchange for 4 $H^+$, with no net charge movement. So the membrane potential does not affect
PMCA's transportation. As for SERCA, it transports 2 $Ca^{2+}$ from cytoplasm to the ER lumen to exchange for 2 $H^+$, but there are no potential difference between the two sides of ER membrane. Therefore the drastic fluctuation of membrane potential during the burst would not affect $Ca^{2+}$'s activity. Thus the quasi-equilibrium assumption($\frac{\mathrm{d}P_{E_1}}{\mathrm{d}t}=\frac{\mathrm{d}P_{E_2}}{\mathrm{d}t}=0$) can be applied to write out $J_{PMCA}$ and $J_{SERCA}$:
\begin{widetext}
\begin{equation}\label{eq:69}
   J_{PMCA}=2\frac{k_5^+k_6^+[ATP]_i[Ca^{2+}]_i^2-k_{5}^-k_6^-[ADP]_i[Pi]_i[Ca^{2+}]_{er}^2}{k_5^-[ADP]_i+k_5^+[Ca^{2+}]_i^2[ATP]_i+k_6^-[Ca]_{er}^2[Pi]_i+k_6^+}
\end{equation}
\begin{equation}\label{eq:70}
    J_{SERCA}=2\frac{k_7^+k_8^+[ATP]_i[Ca^{2+}]_i^2-k_{7}^-k_8^-[ADP]_i[Pi]_i[Ca^{2+}]_{er}^2}{k_7^-[ADP]_i+k_7^+[Ca^{2+}]_i^2[ATP]_i+k_8^-[Ca^{2+}]_{er}^2[Pi]_i+k_8^+}
\end{equation}
\end{widetext}

For the similar reason discussed above, the energy consumed by PMCA and SERCA is ignored, but the thermodynamic consistency is still required:
\begin{equation}\label{eq:75}
\frac{k_5^-k_6^-}{k_5^+k_6^+}=\frac{k_7^+k_8^+}{k_7^-k_8^-}=K_{ATP,hyd}
\end{equation}

The values of all the parameters are listed in the Appendix.

Accordingly, we have ATP hydrolysis energy per unit time consumed by NKA, PMCA and SERCA as:
\[P_{NKA} = -\sigma_{NKA}J_{NKA}\Delta \mu_{NKA} \]
\[P_{PMCA} = -\frac{1}{2}\sigma_{PMCA}J_{PMCA}\Delta \mu_{PMCA} \]
\[P_{SERCA} = -\frac{1}{2}\sigma_{SERCA}J_{SERCA}\Delta \mu_{SERCA} \]
where, \(\Delta\mu_{PMCA} = k_BT\ln(\frac{k_5^+k_6^+[ATP]_i[Ca^{2+}]_i^2}{k_{5}^-k_6^-[ADP]_i[Pi]_i[Ca^{2+}]_{er}^2})\) and \(\Delta\mu_{SERCA} = k_BT\ln(\frac{k_7^+k_8^+[ATP]_i[Ca^{2+}]_i^2}{k_{7}^-k_8^-[ADP]_i[Pi]_i[Ca^{2+}]_{er}^2})\)

If we consider ATP, ATP and Pi as variables instead of fixed cellular environment, we are able to characterize ATP comsumption with time.
\section{\label{sec:level3}Results}

\subsection{The basic properties of sAHP}
\begin{figure*}
\centering
\includegraphics[width=1\linewidth]{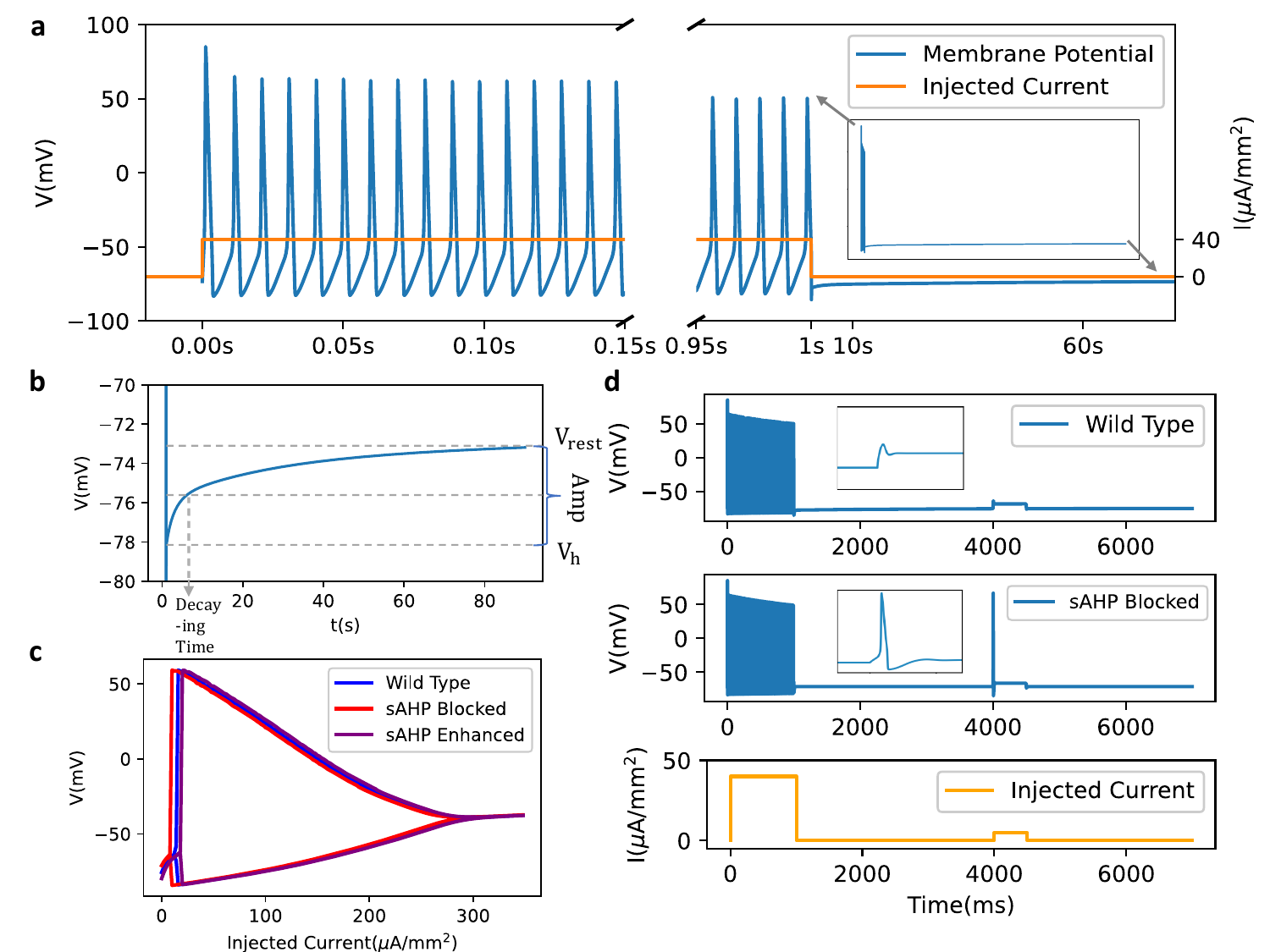}% Here is how to import EPS art
\caption{\label{fig:2} \textbf{The simulation of slow afterhyperpolarization(sAHP) and its function on controlling neural excitability}  \quad\textbf{a.}The simulation of neural bursting and sAHP process. The protocol of simulation is to set $I_{inj} = $40 $\mu$A/mm$^2$ for 1 second to stimulate burst. After shifting $I_{inj}$ to 0, the cell stops spiking and displays a slowly recovering hyperpolarization. Here we used two different time scales to plot the potential fluctuation in the burst and the sAHP, respectively. The curve of the membrane potential change with uniform time scale is plotted in the subfigure of the graph.\ \textbf{b.} Amplification of the sAHP process. We defined the most negative membrane potential after 1.05s (to exclude the influence of \textbf{K(V)} current) as $V_h$, and the resting state membrane potential after long enough time as $V_{rest}$. We denoted $V_{rest} - V_h$ as the AHP amplitude(Amp), and the time for the membrane potential to recover to $\frac{V_{rest}+V_h}{2}$ as the decaying time.\ \textbf{c.} H-H system's bifurcation with Injected Current under normal situation (Wild Type), sAHP Blocked (set $g_{K(Ca)}=0$ and $\sigma_{NKA}=0$), and sAHP Enhanced (double $g_{K(Ca)}$ and $\sigma_{NKA}$) at t=4s in \textbf{a.}. The start point of bifurcation decreases for sAHP Blocked situation and increases for sAHP Enhanced situation.\ \textbf{d.} An specific example of sAHP controlling the neural excitability after a burst. Same with \textbf{a.}We first set $I_{inj}=$ 40 $\mu$A/mm$^2$ for 1 second. Then at t=4s, we simulated a direct current injection with 5 $\mu$A/mm$^2$ density for 0.5 second. A wild type neuron does not response with an action potential, while the one with sAHP Blocked does. It shows that sAHP protects the neuron from continuing to spike after a burst. This simulation is consistent with the experimental observation\cite{Arganda_2007}.}
\end{figure*} 

	A representative pattern of burst and sAHP process is plotted in FIG \ref{fig:2}a. With direct current injected through the electrode, the neuron spikes at a frequency of $\sim$ 200Hz. At t=1s, the injection ceases and the membrane potential quickly hyperpolarizes to a magnitude that is more negative than its bursting and resting state. The potential then goes through a long-lasting hyperpolarization of $\sim$ 80s. This long recovering time distinguishes this process from the afterhyperpolarization after each single spike (also called fast afterhyperpolarization, fAHP). The details of the sAHP process is amplified in FIG \ref{fig:2}b.\par
In FIG \ref{fig:2}b we defined several variables $V_{rest}$, $V_{h}$, Amp, decaying time to describe the behavior of sAHP. It is notable that the hyperpolarization after the burst is actually the sum of the fast AHP and the slow AHP (There are also works proposing the existence of the medium AHP of a moderate time scale between the fAHP and sAHP\cite{Larsson_2013}) Since fAHP is generated by a delay of the K(V) channel turning off(whose relaxation time constant is $\sim$ 0.01s), we defined $V_{h}$ as the most negative membrane potential after 1.05s to exclude the fAHP.\par
	sAHP has been found to affect neuron's excitability after neural bursting. FIG \ref{fig:2}c plotted the amplitude changes of the limit cycle for membrane potential oscillation with the amplitude of the injected current  at t=4s for the bursting protocol in \ref{fig:2}a. The simulation results shows that the neuron goes into the limit cycle oscillation with a lower threshold current amplitude if the pathways generating sAHP is blocked. And the threshold is higher if sAHP is enhanced.(Here we simulate the enhancement by increasing the expression of NKA and K(Ca) channel on the neuron membrane). \par
	FIG \ref{fig:2}d gives a intuitive illustration of the property discussed above. When the after-burst stimulating current is chosen just below the threshold for the Wild Type neuron, the neuron cannot be activated under normal situation, but the neuron with sAHP blocked displays an action potential. The neuron transforms a slight offset of membrane potential to a '0' or '1' difference in the neural coding by making use of the phase transition behavior of the neuron's excitable system. Similar results can be found in many experiments\cite{Arganda_2007}\cite{Zhang_2012}\cite{Zhang_2015}.\par
	This transformation seems to be under strict restriction since it only happens when the injected current density lies just within the bifurcation point change, which is a small range of only several $\mu$A/mm$^2$. However, the brain is considered to usually work on the critical point\cite{Shew_2012}\cite{Mu_oz_2018}. It is theoretically proved that the criticality in the neural network optimizes the network's sensitivity, dynamical range and correlation length, etc\cite{Langton_1990}\cite{Boedecker_2011}. Therefore, maybe it is common for the synaptic input current to lie in the range required for sAHP to come into play.
\begin{figure*}
\centering
\includegraphics[width=1\linewidth]{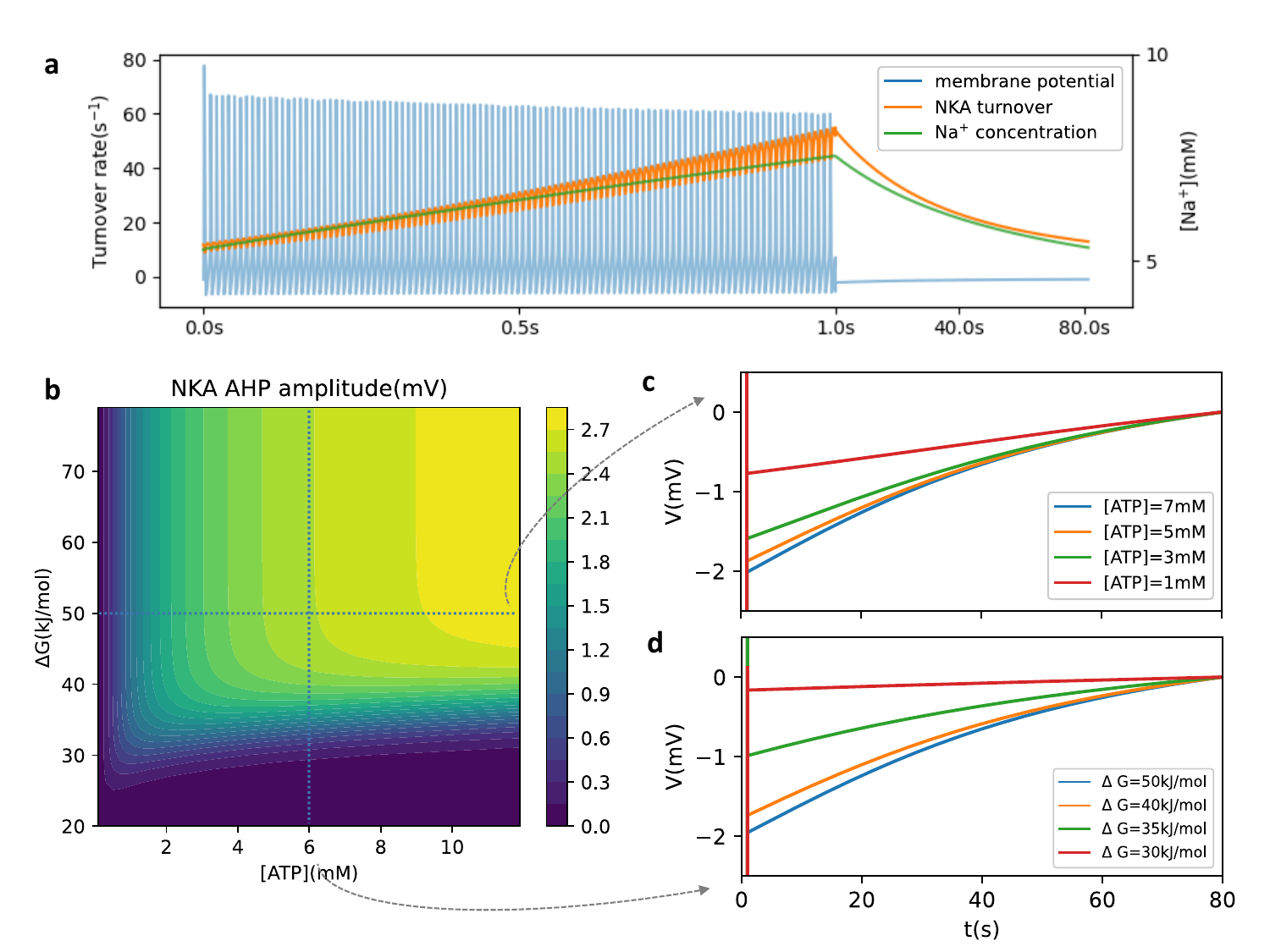}% Here is how to import EPS art
\caption{\label{fig:3} \textbf{$Na^+$-$K^+$ ATPase’s activity in AHP generation and its dependence on ATP concentration and the phosphorylation energy $\Delta$G.(With K(Ca) blocked)}\quad\textbf{a.} NKA’s turnover rate, $Na^+$ concentration and membrane potential's change along with time during a burst-sAHP process. The energy level is set under physiological conditions: [ATP]=6mM and  $\Delta$G=50kJ/mol. Here we use different time scale during the burst and the sAHP for the same reason in Fig. 2.\ \textbf{b.} sAHP's amplitude(Amp)'s dependence on the energy supply conditions. Insufficient ATP concentration or small $\Delta$G will decrease the sAHP amplitude.\ \textbf{c.}The curve of membrane potential changes during the sAHP under different ATP concentration, with $\Delta$G fixed to be 50kJ/mol.Different energy supply conditions will not only affect the sAHP process but also change the resting potential. In order to clearly compare the Amp in different condition, we actually plotted $V-V_{rest}$’s change along with time here. We applied this technique in both FIG 3c,d and FIG.4d,e.\ \textbf{d.}The curve of membrane potential changes during the sAHP under different $\Delta$G levels, with [ATP] fixed to be 6mM.}
\end{figure*}
\subsection{Energy Dependence of the sAHP}
Since the two pathways to generate sAHP are both linked with ATP-hydrolysis ATPases, the change of energy supply condition may affect sAHP's behavior. Based on the sensitivity of sAHP's modulation dicussed in the last section, it is reasonable to guess that the ATP concentration or $\Delta$ G level in the cell can affect neural excitability through changing sAHP. With our model we can easily shut down one pathway and discuss the energy dependence of the other. The simulation results shows that the energy dependent behaviors of the sAHP generated by these two different pathways are different.
\begin{figure*}
\centering
\includegraphics[width=1\linewidth]{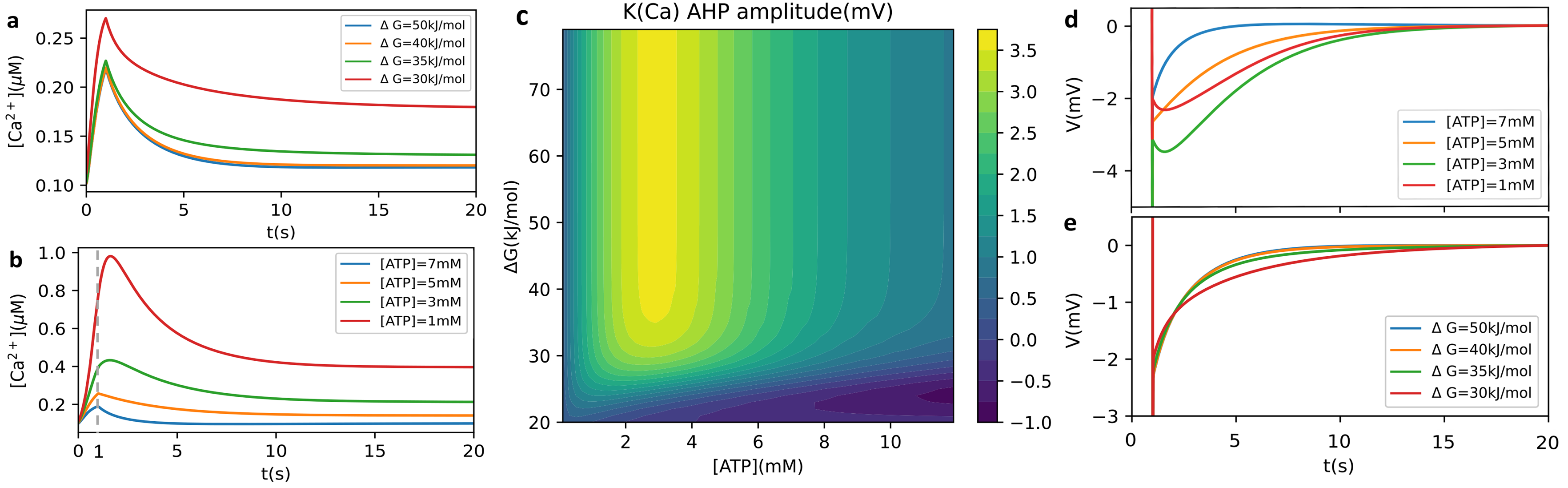}% Here is how to import EPS art
\caption{\label{fig:4} \textbf{K(Ca) channel’s contribution in AHP generation and its dependence on ATP concentration and the phosphorylation energy $\Delta$G.(With NKA blocked)}\quad\textbf{a.}The changes of the cytoplasm $Ca^{2+}$ concentration during the bursting and the AHP period under different ATP concentration, with $\Delta$G fixed to be 50kJ/mol.\ \textbf{b.}The changes of the cytoplasm $Ca^{2+}$ concentration during the bursting and the AHP period under different $\Delta$G levels with [ATP] fixed to be 6mM.\ \textbf{c.}sAHP's amplitude(Amp)'s dependence on the ATP concentration and $\Delta$G level. As the ATP concentration changing from 1 mM to 7 mM, the AHP amplitude is first increased and then decreased. While the $\Delta$G level’s increase cause monotonous decrease of AHP amplitude.\ \textbf{d.}The curve of membrane potential changes during the sAHP under different ATP concentration, with $\Delta$G fixed to be 50kJ/mol.\ \textbf{e.}The curve of membrane potential changes during the sAHP under different $\Delta$G levels, with [ATP] fixed to be 6mM.}
\end{figure*}

\subsubsection{Energy Dependence of the NKA-Modulating sAHP Part}

The $Na^+$-$K^+$ ATPase(NKA) keeps the balance of intracellular $Na^+$ and $K^+$ by active transportation, so it uses  phosphorylation energy $\Delta$G to overcome the electrical and chemical gradient through the cell membrane. Therefore, the change of the membrane potential and the change of intracellular and extracellular $Na^+$ and $K^+$ concentration will both affect the function of the NKA. FIG \ref{fig:3}a shows the change of NKA's turnover rate, $Na^+$ concentration and neural membrane potential along with time during a burst-sAHP process. During the burst, the turnover rate oscillates with membrane potential's drastic variation. In the meantime, $Na^+$ flows into the neuron and $K^+$ flows out through passive transportation of \textbf{Na(V)} and \textbf{K(V)}, which results in a decreased chemical gradient through the membrane. It will raise the NKA’s turnover rate. FIG \ref{fig:3}a shows the increase of the turnover rate of $Na^+$-$K^+$ ATPases along with the change of the $Na^+$ concentration.\par
The rate reaches its peak at the end of the burst period. At that moment, the injected current is removed, \textbf{Na(V)} and \textbf{K(V)} channels are closed, leaving the outward current of NKA the dominant part to change the membrane potential. So, the membrane is hyperpolarized to its most negative potential by the NKA current. \par
The NKA continues to remove redundant ions after the burst, adjusting the concentration of $Na^+$ and $K^+$ back to its normal level, which in turn decreases the turnover rate of NKA itself. Therefore, the outward current gets smaller and the membrane gradually recovered from hyperpolarization.\par
Since the rotation of NKA is triggered by ATP hydrolysis, changing the energy supply condition will definitely affect its efficiency. With an  insufficient energy supply, the reduced turnover rate will cause smaller outward current and then affect the sAHP amplitude. FIG \ref{fig:3}b simulate the sAHP amplitude under different ATP concentrations or $\Delta$G levels. It shows that insufficient ATP concentration or $\Delta$G will both decrease the amplitude of the sAHP generated by NKA. The ATP-$\Delta$G phase graph shows the lower boundary for sAHP generation and also a saturation at high ATP concentration or $\Delta$G level. FIG \ref{fig:3}c,d plotted specific examples of sAHP under different energy conditions.

\subsubsection{Energy Dependence of the K(Ca)-Modulating sAHP Part}
\begin{figure*}
\centering
\includegraphics[width=1\linewidth]{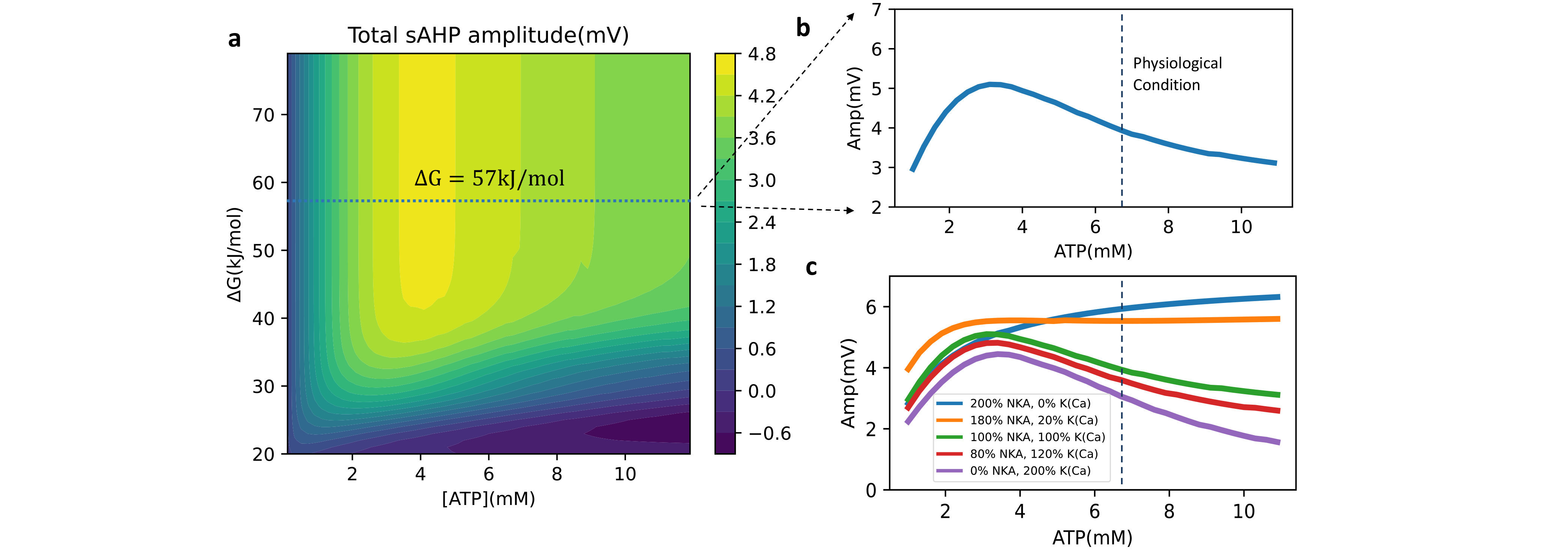}% Here is how to import EPS art
\caption{\label{fig:5}  \textbf{The combined effect of NKA and K(Ca)'s modulation to the AHP amplitude under different energy supply conditions.}\quad\textbf{a.}The sAHP's amplitude changes under different ATP concentrations and $\Delta$G levels.\ \textbf{b.}The change of Amp along with ATP concentration when $\Delta$G is set at 57kJ/mol.\ \textbf{c.}The change of Amp along with ATP concentration when $\Delta$G is set at 57kJ/mol under different proportions of NKA and K(Ca)'s expression.}
\end{figure*}
The K(Ca) channel is the other component in the model that modulate the sAHP. This channel is activated by the $Ca^{2+}$ and generate an outward $K^+$ current. The concentration of $Ca^{2+}$ affects the opening probability P of the channel through a 2-nd Hill equation manner:P$\propto \frac{[Ca^{2+}]^2}{[Ca^{2+}]^2+k_Ca^2 }$.\par
During the burst, the $Ca^{2+}$ level in the cytoplasm goes through distinct changes. As stated in the previous section\ref{sec:level2}, there are mainly two sources for the cytoplasm $Ca^{2+}$: the extracellular fluid and the calcium store inside the cell (mainly inside the endoplasmic reticulum, ER), both have a much higher $Ca^{2+}$ concentration than the cytoplasm. \par
Similarly, the energy supply conditions affect $Ca^{2+}$ concentration changes since there are ATP-hydrolysis-triggered $Ca^{2+}$ ATPases get involved. FIG\ref{fig:4}a,b shows the curves of $Ca^{2+}$ concentration changes in burst-sAHP process when ATP concentration or $\Delta$ G level is changed.\par
The depolarization of membrane potential during the burst opens the Ca(V) channel and causes  $Ca^{2+}$ influx. The rise of $Ca^{2+}$ level in the cytoplasm not only raises the turnover rate of $Ca^{2+}$ATPases on the cell and ER membrane (SERCA, PMCA) in a similar way as the increase of [$Na^+$] raises the $Na^{+}$-$K^{+}$ ATPases's activity, but it also triggers the opening of the IP$_3$R channel on the ER membrane. Therefore, the $Ca^{2+}$inside the ER has the way to flow out into the cytoplasm and launches a positive feedback loop. So the cytoplasm $Ca^{2+}$ concentration continues to increase until the activities of  $Ca^{2+}$ATPases rise to a high enough level to remove the $Ca^{2+}$ into the ER or out of the cell. \par
When the burst is ceased and the membrane depolarization is stopped, the Ca(V) channel closes but the IP$_3$R channel, not affected by membrane potential, does not; so if the flux through IP$_3$R is larger than the summed flux through PMCA and SERCA, the $Ca^{2+}$ concentration will continue to rise until the ATPases' flux exceeds IP$_3$R's. That is why the peak of the $Ca^{2+}$ concentration appear after the burst ends when [ATP]=1mM, 3 mM in \ref{fig:4}a. The simulation of $Ca^{2+}$ concentration's changing behavior is similar with the experiment results\cite{Gulledge_2013}.\par
As showed in FIG\ref{fig:4}a,b, lower ATP or $\Delta$G level will result in $Ca^{2+}$ ATPases’ inefficiency, so more $Ca^{2+}$ accumulates inside the cytoplasm. It leads to a higher $Ca^{2+}$ level peak and also a higher equilibrium $Ca^{2+}$ concentration. A higher $Ca^{2+}$ level peak triggers a larger outward \textbf{K(Ca)} current maximum, while a higher equilibrium $Ca^{2+}$ concentration enlarge the \textbf{K(Ca)} current at rest. The sAHP is generated by the difference between the two, so Amp's energy dependence is determined by how the difference between maximal and resting  \textbf{K(Ca)} current change with energy supply. The simulation in \ref{fig:4}c shows that as the ATP concentration changing from 1 mM to 7 mM, the AHP amplitude first increases and then decreases. And the $\Delta$G level’s increase cause monotonous decrease of AHP amplitude. FIG \ref{fig:4}d,e plotted specific examples of sAHP under different energy conditions.

\subsection{The Combining Effect of NKA and K(Ca) on the Total AHP’s Energy Dependence}

Based on the discussions above, we then use the model to simulate how the total sAHP amplitude's energy dependence, which is plotted in the ATP-$\Delta$G phase graph FIG\ref{fig:5}a. Comparing with the graph with one pathway shutting down (FIG\ref{fig:3}b and FIG\ref{fig:4}c), the contour plot appears to be more flat, which implies the combining modulations of NKA and K(Ca) are able to maintain a relatively stable sAHP amplitudes when ATP concentration or $\Delta$G fluctuates and in turn keep the stability of sAHP's function of adjusting neural excitability.\par
To make it clearer, we can set $\Delta$G to be the physiological level (57kJ/mol) since FIG\ref{fig:3}b, FIG\ref{fig:4}c, FIG\ref{fig:5}a all show that ATP concentration has a larger effect on Amp, and plotted the change of Amp with ATP concentration changes, which is plotted in FIG\ref{fig:5}b. We can then use the amplitude of Amp variation under certain range of ATP concentration changes to measure the ability to maintaining stability of this combining modulation. We found that this ability can be adjusted through changing the proportion of the amount of NKA and K(Ca)'s expression on the neural membrane, as show in FIG\ref{fig:5}c. In addition, changing the proportion of the amount of NKA and K(Ca)'s expression can also qualitatively alter the behavior of sAHP's energy dependence: when NKA's amount dominant, Amp decreases with lower ATP concentration, while the K(Ca) dominating situation is just the opposite.

\subsection{Mimicking Hippocampal CA1 Pyramidal Neurons of Aging Animals by Reduced Intracellular ATP Concentrations and Upregulating Membrane Calcium Channels
}
\begin{figure*}
\centering
\includegraphics[width=1\linewidth]{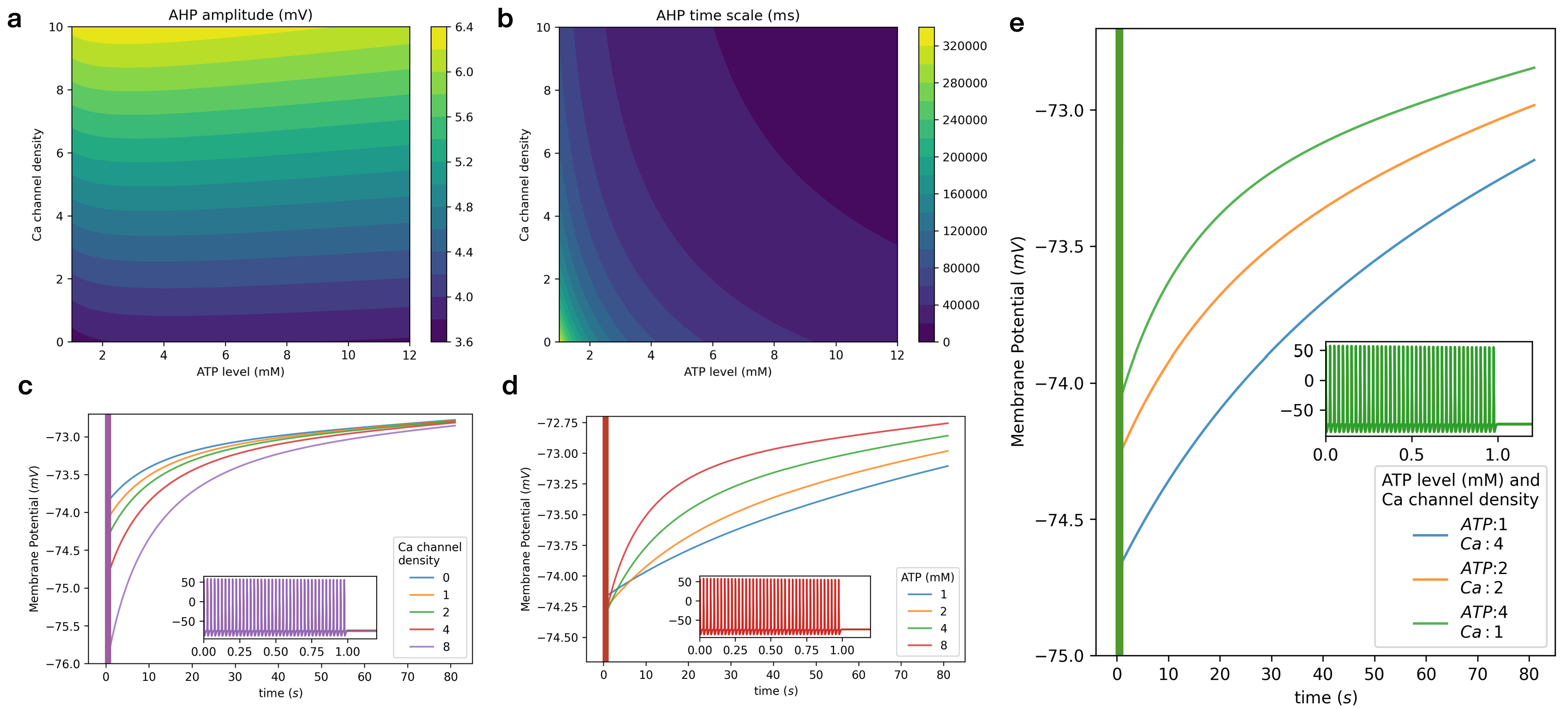}% Here is how to import EPS art
\caption{\label{fig:6}  \textbf{The combined effect of plasma membrane L-type voltage-gated
calcium channels (LGVCC) and cellular ATP level to the AHP amplitude and AHP  time scale.}\quad\textbf{a.}The sAHP's amplitude changes under different calcium channel density and ATP concentrations.\ \textbf{b.}The sAHP's time scale changes under different calcium channel density and ATP concentrations.\ \textbf{c.}The change of AHP along with calcium channel density when ATP is set at 4mM.\ \textbf{d.}The change of AHP along with ATP concentration when calcium channel density is set at 2. \textbf{e.}The change in AHP with simultaneous decrease in ATP concentration and increase in Ca ion channel density. {\textbf{a-e}.}$\Delta$G is set at 57kJ/mol. {\textbf{c-e}.}The illustration depicts the dynamic behaviour of the membrane potential in response to external stimulation at 40 Hz.}
\end{figure*}

To verify the effects of intracellular ATP levels and the free energy of ATP hydrolysis on neuronal processes, which represents level of energy metabolism, on sAHP, we conducted a systematic literature review and analysis. The most direct biological process associated with the decline in neuronal energy metabolism is aging. The reduction in energy metabolism within the nervous system during aging has been confirmed at multiple levels, including imaging evidence of decreased cerebral blood flow, oxygen consumption, and glucose utilization\cite{https://doi.org/10.15252/embj.201695810}\cite{MATTSON20181176}\cite{Li2023}, as well as direct measurements showing reduced intracellular ATP concentrations\cite{traut1994physiological}\cite{xiong2023vimar}. Furthermore, we noticed compelling evidence indicating abnormalities in cellular calcium homeostasis during aging, characterized by an upregulation of plasma membrane L-type voltage-gated calcium channels (LGVCC) \cite{thibault1996increase} and an increase in baseline intracellular calcium levels\cite{xiong2023vimar}. These characteristics make aging an ideal biological reference for our simulation results. In this section, we focused on exploring how intracellular ATP level and $Ca$ ion channels density affect sAHP. To better mimic actual neuronal processes under physiological conditions, $\Delta$G is set at 57kJ/mol and a fixed externally applied stimulation frequency was used, set at 40 Hz to simulate neuronal activity during the thinking period.

As shown in Fig\ref{fig:6}a, Ca-ion channel density appears to be positively associated with AHP amplitude, as a greater density of Ca-ion channels causes a greater influx of $Ca^{2+}$ during an action potential sequence, which will then cause $K(Ca)$ channel activation, resulting in a more intense hyperpolarisation. However, in this case, ATP did not have a significant effect on the amplitude of AHP, mainly because the kinetic behaviour during this time was more maintained by the original ion concentration. On the other hand, as shown in Fig\ref{fig:6}b, ATP levels significantly affect the time scale of the AHP, with higher ATP shortening the time scale of the AHP and restoring it to the resting potential more quickly. the density of Ca ion channels also affects, but to a relatively small extent. It can also be seen from the dynamics of membrane potentials in Fig\ref{fig:6}c that membrane potentials were significantly more hyperpolarised when Ca channel density increased after 1s of action potentials elicited by 40 Hz external stimulus. And as shown in Fig\ref{fig:6}d, a higher ATP concentration helps to restores membrane potential more quickly.

Our simulation results demonstrated relative consistency with prior biological experimental data. In hippocampal CA1 pyramidal neurons of aging animals—which are the most classic model cells for sAHP—significant electrophysiological changes were observed compared to neurons from younger animals. These changes include an increased sAHP amplitude and a prolonged membrane potential recovery duration\cite{matthews2009fast}\cite{tombaugh2005slow}. For comparison, simulation results are shown in Fig\ref{fig:6}e that under conditions of reduced ATP concentration and increased LGVCC density, the sAHP amplitude significantly increases, and its time scale is notably prolonged. These findings corroborate the existing biological experimental results.

\section{\label{sec:level4}Conclusion and Discussion}
In this paper, we constructed a biophysically-based model to describe the generation of the slow afterhyperpolarization after the neural bust. The model contains two pathways to generate sAHP: the $Na^+$-$K^+$ ATPase and the K(Ca) channel, both involve ATP-hydrolysis ATPases and therefore are modulated by the intracellular energy level. We simulated the effects of ATP concentration and $\Delta$ G level's variations on sAHP's amplitudes and pointed out that NKA and K(Ca)'s combining modulation of sAHP amplitude display a trade-off to maintain a relative stable sAHP amplitude when the energy condition inside the neuron fluctuates. This property may have significance in understanding neuron's maintenance of homeostasis. However, the prediction has to be tested through more experiments. A more precise experimental validation may be based on an in vitro cell-based system that can simultaneously modifies intracellular ATP and phosphorylated free energy levels.\par
The energetic property of neural behavior is always a focus of brain research, especially research about some neural degenerative disease. Because the underlying mechanisms of these disease are often considered to be related with the reduced level of energy supply due to several aging effects such as the malfunction of the mitochondria. Further, understanding such mechanism may generate therapies to resist the effect of insufficient energy supply. \par
The sAHP phenomenon has been found some aging effect that may relate with energy supply. It is observed that the sAHP amplitude in the aged rats's neuron is enlarged\cite{Hemond_2005}. If we assume that there is a energy level cutting down related with aging, our model can reproduce this phenomenon with a K(Ca) dominating sAHP.\par
Besides, our model shows that changing the proportion of NKA and K(Ca)'s expression quantities can alter sAHP's modulating behavior on the neural excitability when energy level changes: by raise the level of K(Ca) expression, the sAHP may shift from increasing neural excitability to suppressing it when ATP level decreases, for example. Actually, the change of K(Ca) channels expression level in aged neuron has already been proved to have link with the recession of LTP and cognitive functions\cite{Blank_2003}\cite{Wolfart_2001} and man-made overexpression of K(Ca) channels was found resulting in deficit on intrinsic pacemaking precision in the neural work\cite{Martin_2016}. But the underlying mechanism is still unclear.\par 
Furthermore, model indicated that a reduction in ATP levels and an increase in $Ca$ ion channel density, two phenomena observed in neurons of aged mice, also influence sAHP behaviour, exhibiting sAHP behaviour comparable to that observed in mice exhibiting cognitive decline. In addition to the electrophysiological findings, we noted that aging animals exhibit significantly poorer performance in hippocampus-dependent learning tasks compared to younger animals. This decline in learning ability has been shown to have a causal relationship with the sAHP changes observed in hippocampal neurons. Importantly, this sAHP-mediated cognitive decline is cross-species, including tasks such as eyeblink conditioning in rabbits\cite{moyer2000increased} and the Morris water maze in rodents\cite{tombaugh2005slow}. This highlights the fundamental significance of sAHP in neural computation. Several studies on the role of sAHP in neuronal information processing support our perspective\cite{higgs2006diversity}\cite{doi2023spontaneous}. Given that sAHP plays a critical role in regulating burst firing ,which is a primary spiking mode for neuronal information processing, we propose a hypothesis that sAHP abnormalities caused by impaired energy metabolism mediate a reduction in neuronal information bandwidth, thereby contributing to non-pathological cognitive decline in aging individuals. In order to characterise cognitive abilities, we propose the metric known as the information flux potential (IFP) of neurons, which is defined as the number of action potentials that can be processed per unit time. Our findings indicate a strong correlation between information flux potential and $Ca$ ion channel density (see Appendix). Future research will focus on the capacity of individual neurons to process signals. So would the sAHP's energy dependence be the missing link of these experimental results? More theoretical and experimental analyses are needed.

% \begin{acknowledgments}

% \end{acknowledgments}

\nocite{*}

% \bibliography{ref}% Produces the bibliography via BibTeX.
%apsrev4-2.bst 2019-01-14 (MD) hand-edited version of apsrev4-1.bst
%Control: key (0)
%Control: author (72) initials jnrlst
%Control: editor formatted (1) identically to author
%Control: production of article title (-1) disabled
%Control: page (0) single
%Control: year (1) truncated
%Control: production of eprint (0) enabled
%

% \end{CJK}
\end{document}